\pdfoutput=1

\documentclass[11pt]{article}

\usepackage[final]{acl}

\usepackage{times}
\usepackage{latexsym}

\usepackage[T1]{fontenc}

\usepackage[utf8]{inputenc}

\usepackage{microtype}

\usepackage{inconsolata}

\usepackage{graphicx}

\usepackage{enumitem}

\title{Oversight Structures for Agentic AI in Public-Sector Organizations}

\author{
\textbf{Chris Schmitz\textsuperscript{1}},
\textbf{Jonathan Rystrøm\textsuperscript{2}},
\textbf{Jan Batzner\textsuperscript{3,4}}
\\
\textsuperscript{1}Centre for Digital Governance, Hertie School, Germany,\\
\textsuperscript{2}Oxford Internet Institute, University of Oxford, UK,\\
\textsuperscript{3}Weizenbaum Institute Berlin, Germany,\\
\textsuperscript{4}Technical University Munich, Germany
\\
\small{
\textbf{Correspondence:} \href{mailto:ch.schmitz@hertie-school.org}{ch.schmitz@hertie-school.org}
}
}

\begin{document}
\maketitle
\begin{abstract}
This paper finds that the introduction of agentic AI systems intensifies existing challenges to traditional public sector oversight mechanisms --- which rely on siloed compliance units and episodic approvals rather than continuous, integrated supervision. We identify five governance dimensions essential for responsible agent deployment: cross-departmental implementation, comprehensive evaluation, enhanced security protocols, operational visibility, and systematic auditing. We evaluate the capacity of existing oversight structures to meet these challenges, via a mixed-methods approach consisting of a literature review and interviews with civil servants in AI-related roles. We find that agent oversight poses intensified versions of three existing governance challenges: continuous oversight, deeper integration of governance and operational capabilities, and interdepartmental coordination. We propose approaches that both adapt institutional structures and design agent oversight compatible with public sector constraints.
\end{abstract}

\section{Introduction} \label{sec:intro}
Artificial Intelligence (AI) technologies, particularly Large Language Model-based agents, hold the potential to fundamentally transform Public Sector Organizations (PSOs) via efficiency enhancements and process automation  \citep{ilves2025agentic, straubAIBureaucraticProductivity2024, shavitPracticesGoverningAgentic2023}. However, PSOs often lack the institutional capacity and agile governance\footnote{Throughout this article, we use the term ``governance'' to refer specifically to tasks and structures for oversight of LLM agents and other digital projects, rather than to the political-scientific definition, which may encompass all PSO activities.} structures required for responsible adoption \citep{lawrenceBureaucraticChallengeAI2023}. These challenges are driven by characteristic features of bureaucratic organizations, such as procedural rigidity, hierarchical accountability, and dual demands for transparency and effectiveness, which complicate both AI deployment and oversight \citep{madanAIAdoptionDiffusion2023,neumannExploringArtificialIntelligence2024a}.

\begin{figure}[t]
    \centering
    \includegraphics[width=0.97\linewidth]{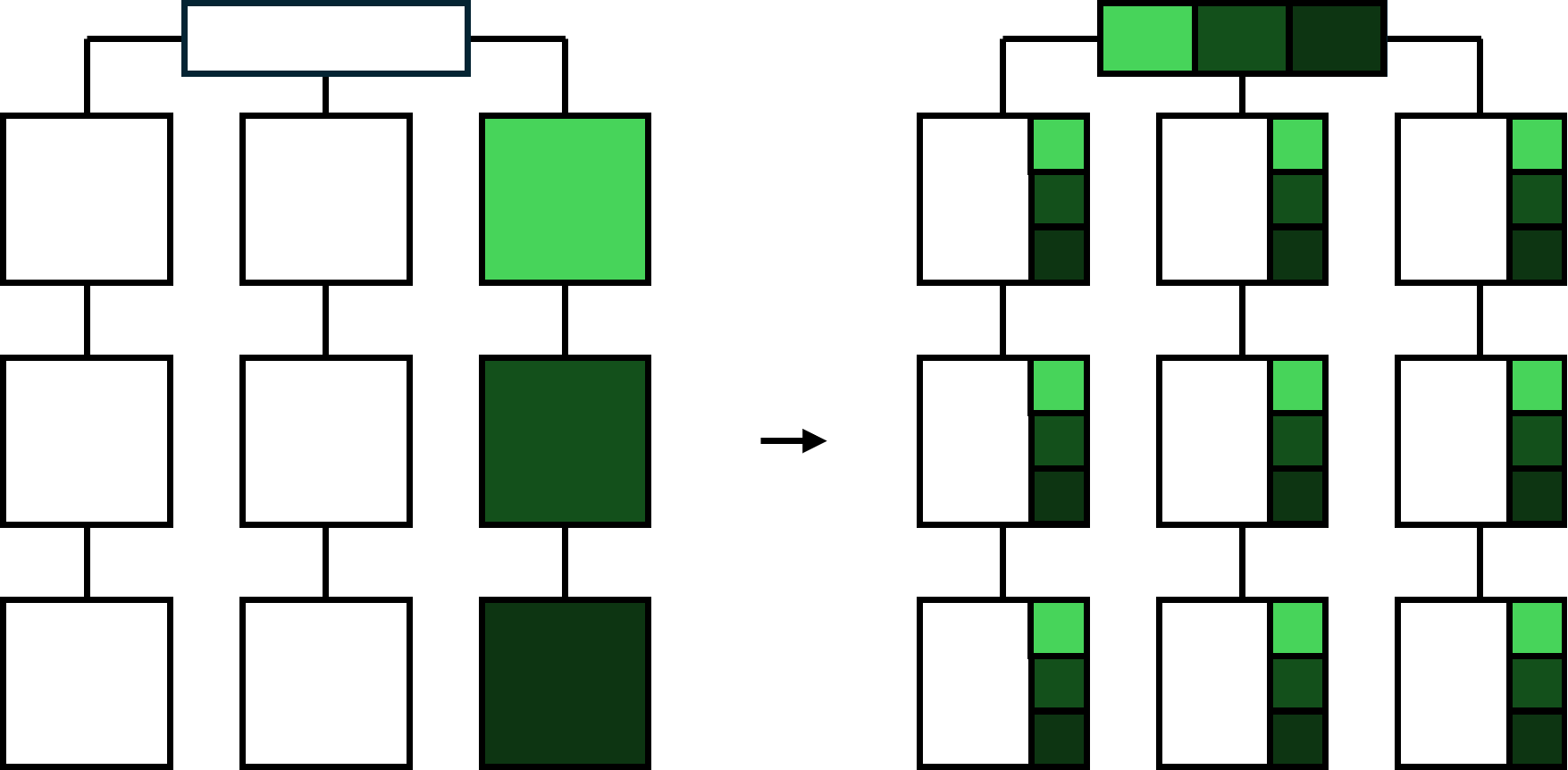}
    \caption{To enable agent oversight, PSOs may move from standalone digital governance functions (in green) towards centrally coordinated, distributed governance.}
    \label{fig:governance-structure}
\end{figure}

Although recent work has proposed best practices for AI agent governance \citep{shavitPracticesGoverningAgentic2023, chanVisibilityAIAgents2024}, it therefore remains unclear whether PSOs are structurally and procedurally equipped to implement these recommendations. The governance of agent systems likely requires both technical implementation and institutional adaptation --- affecting processes, oversight mechanisms, and digital infrastructure \citep{madanAIAdoptionDiffusion2023,neumannExploringArtificialIntelligence2024a}. We ask the following research question:
\\\\
\textit{How do existing public sector AI governance structures meet the requirements for responsible agent deployment?}
\\\\
Based on interviews with civil servants, we find that agent governance requirements exasperate existing digital governance challenges in PSOs. We first outline the potential of AI agents in the public sector (§\ref{sec:agent-potential}) and emerging agent governance tasks (§\ref{sec:gov-agents}) . We derive a hypothesis on challenges these tasks pose to public-sector oversight structures from public administration literature (§\ref{sec:theory}), and describe our interview methodology (§\ref{sec:methods}). We summarize our findings on existing oversight structures (§\ref{sec:governance-pso})  and, finally, chart the ways agent oversight intensifies existing governance challenges in PSOs (§\ref{sec:map-challenge}). Our core contribution is an evaluation of the feasibility of agent oversight in current real-world public-sector contexts.

\section{Potential of Agents in PSOs} \label{sec:agent-potential}
Recent work demonstrates the application of LLM-based agents to tasks in narrow domains. Examples include implementations of agents designed to emulate domain experts such as academic researchers \cite{schmidgall2025agentlaboratoryusingllm}, office workers \cite{gurRealworldWebAgentPlanning2023}, developers \cite{jimenezSWEbenchCanLanguage2023} or healthcare professionals \cite{zhao-nurses}. Moreover, major LLM providers started introducing their Computer Use Agents for end-users in fall 2024 \cite{anthropic2024computer,openai2025operator,pichai2024gemini}.

While the AI community acknowledges the potential of such increasingly agentic systems \cite{liuAgentBenchEvaluatingLLMs2023,lu-etal-2025-toolsandbox}, which integrate autonomous learning, tool use, and complex reasoning, \citet{kapoorAIAgentsThat2024} emphasize their substantial current limitations, including susceptibility to benchmark overfitting, prohibitive operational costs, and unnecessarily complex computational architectures --- which each scale with domain generality. Narrow, domain-specific agentic systems may therefore represent a more practical and immediately valuable development pathway. 

The public sector is such a domain, as public-sector tasks possess ideal characteristics for automation: they are comprehensively documented, exhibit consistent structural patterns, and involve high-frequency repetition \cite{bullock_2020_ArtificialIntelligenceBureaucraticformdiscretion}. Focusing agent development on these well-structured tasks provides a pragmatic pathway toward realizing the benefits of AI agents while mitigating their current limitations.

Previous waves of digitization standardized data formats, process explanations, and decision frameworks \cite{bovens_street-level_2002}, contributing to \textit{bureaucratic rationality}: the formal legibility and transparency of bureaucracies, from which they gain their legitimacy. LLM-based agents potentially continue this process, e.g., by supporting documentation requirements in specific contexts \cite{newmanDigitalTechnologiesArtificial2022, mokanderArtificialIntelligenceRationalization2024a}. 

Conversely, due to their well-known mechanistic inexplicability \cite{sharkeyOpenProblemsMechanistic2025}, AI systems including agents may also hinder, rather than enhance, the further rationalization of state entities. Given the fundamental importance of correctly-attributed responsibility to bureaucratic legitimacy \cite{cetinapresuelAdoptionArtificialIntelligence2024}, successful integration of agentic systems therefore depends on the careful design of accountability mechanisms.

The realization of this potential further depends on the resolution of myriad practical challenges commonly faced in PSOs: resource and capacity strain, varying levels of digitalization of existing processes, and growing internal and external reporting requirements, to name a few \cite{lawrenceBureaucraticChallengeAI2023}.

\section{Agent Governance Challenges} \label{sec:gov-agents}
Deploying LLM-based agents in the public sector presents novel governance challenges over and above those introduced by non-agentic software. In this section, we identify five interdependent governance areas that collectively enable responsible agent deployment. 

\paragraph{Distributed Implementation:}
Compared to traditional IT projects, LLM agent-based systems add layers of complexity due to their many interacting components \citep{wangSurveyLargeLanguage2024}, which often span organizational units. A crucial challenge is to appropriately assign responsibility for different parts of the systems and ensure well-functioning cross-unit communication channels. For instance, implementing a customer query agent requires coordination between IT, customer service, and business intelligence units \citep[for data access;][]{romeAskMeAnything2024}.  

\paragraph{Visibility:}
Visibility into agent activity is an essential prerequisite for both organizational awareness \citep{straubArtificialIntelligenceGovernment2023} and the operational capabilities to identify and intervene in misbehaving systems \citep{chanVisibilityAIAgents2024}. Visibility is generated at both the system level,  via agent indexes similar to model cards \citep{mitchellModelCardsModel2019,derczynskiAssessingLanguageModel2023}, and the operational level, via real-time monitoring \citep{chanVisibilityAIAgents2024}. Such monitoring enables human-in-the-loop safeguards for continuous oversight and proper attribution of responsibility \citep{anannySeeingKnowingLimitations2018}.

\paragraph{Evaluation:}
Pre-deployment evaluation ensures that agent systems work as intended \citep{wangSurveyLargeLanguage2024}. These evaluations must include system-level, task-specific, and comprehensive metrics \citep[e.g.,][]{jimenezSWEbenchCanLanguage2023,kapoorAIAgentsThat2024}, rather than general, component-level evaluations \citep[e.g.,][]{yeToolSwordUnveilingSafety2024}. Evaluation should enable socio-technical comparisons with existing human-based systems, and therefore should be designed in close collaboration with operational workers \citep{selbstFairnessAbstractionSociotechnical2019}.

\paragraph{Security:}
AI Agents present novel security challenges due to their complexity \citep{dengAIAgentsThreat2025}. Threats include jailbreaking \citep{tianEvilGeniusesDelving2024}, data exfiltration \citep{zengGoodBadExploring2024}, and exposure to DDoS attacks \citep{zhangBreakingAgentsCompromising2024}. While many of these challenges fit well under existing cybersecurity frameworks \cite{krumayEvaluationCybersecurityManagement2018} and cybersecurity hygiene practices \cite{guptaCybersecurityHygieneCyber2022}, the autonomy of Agents might require novel governance strategies and practices \cite{dengAIAgentsThreat2025}. Addressing these challenges requires cross-departmental threat mapping, mitigation strategies, and red-teaming efforts \citep{inieSummonDemonBind2025}. 

\paragraph{Auditing:}
Agent auditing capacity should build on the previous capacities to ensure holistic compliance across processes, components, and applications \citep{mokanderAuditingLargeLanguage2024}. Agent auditing provides external compliance validation which ensures that efforts within Visibility, Evaluation, and Security are up to proper standards \cite{sandvigAuditingAlgorithmsResearch2014}. While no formalized auditing frameworks for agents are well-established \citep{chanHarmsIncreasinglyAgentic2023}, PSOs should nonetheless strive to implement both entities and standards for independent auditing procedures.

\section{Public Administration Context} \label{sec:theory}

This section draws on public administration theory to hypothesize that governing LLM-based agents challenges existing governance structures in PSOs.

PSOs are overwhelmingly Weberian bureaucracies \cite{weber_1991_MaxWeberEssaysSociology} with strict delineation of responsibilities, specialization, hierarchical structure, and well-documented, impersonal processes \cite{clegg_weber_2009}. Tasks are differentiated into secluded units \cite{blau_formal_1970} with limited, formalized intra-unit communication, in the case of governance often in the form of approval processes. Governance structures for AI, where existent, often inherit these features, characterized by episodic approvals and externalization  \cite{lawrenceBureaucraticChallengeAI2023}. Bureaucracies may arrive at these structures via ``isomorphic'' processes, in which they mimic similar organizations without considering whether these structures are well-suited for their individual case \cite{dimaggio_iron_1983}.

The well-documented inefficiencies of these structures \cite{niskanen_bureaucracy_1971} have led to the development of several more dynamic governance frameworks, notably New Public Management \cite{hood_public_1991}, Digital-Era Governance \cite{10.1093/acprof:oso/9780199296194.001.0001}, ``governing by network''  \cite{goldsmith_governing_2004} and derived meta-governance approaches \cite{sorensen_metagoverning_2017}. Each of these have found limited success in practice, in part due to the persistence of external factors forcing rationalization, such as administrative law and freedom of information regulation \cite{pozen_transparencys_2018}.

We therefore hypothesize that PSOs' existing governance structures are only partially compatible with the requirements posed by the integration of LLM-based agents. While some novel agent governance tasks may be well-situated within existing bureaucratic oversight structures, holistically governing LLM-based agents will require adaptations that challenge the typical segmentation, timing, and location of governance activities within PSOs. 

\section{Interview Methodology} \label{sec:methods}
In addition to our literature review, we conducted six semi-structured interviews with German officials across six agencies: three federal, two state, and one municipal agency. All participants were recruited through a certificate course on AI in the public sector coordinated by the first author. The interviews followed the interview guide in Appendix \ref{app:interview-guide}. We qualitatively analyzed the interviews using open coding \citep{charmazConstructingGroundedTheory2006}. Finally, we present the COREQ checklist in Appendix \ref{app:coreq} \cite{tongConsolidatedCriteriaReporting2007} following best practices in the field \citep{adeoye-olatundeResearchScholarlyMethods2021}.

\section{Findings on Existing Oversight Structures} \label{sec:governance-pso}

Here we summarize our interview findings into key attributes of existing governance structures for digital projects within PSOs.

\paragraph{Legal Motivation:} Well-established governance generally only exists where legally mandated. In Germany, AI systems as of today underlie no specific laws. As a result, in the surveyed organizations AI governance is largely handled through structures established for traditional digital and process governance. Few organizations have AI-specific governance, such as logging and oversight requirements, in place; only one interviewee cited an AI-specific governance body with formal authority in their organization.

\paragraph{Dedicated Governance Units:} Established digital governance responsibilities include data protection, cybersecurity, and accessibility. In the studied organizations, each of these are held by dedicated units or individuals, which frequently sit in standalone positions across the organization --- in the case of data protection, this is legally required in the EU. These individuals wield significant authority, including veto rights on projects. 

\paragraph{Event-triggered Involvement:} Interviewees in implementing departments report \textit{proactively} consulting governance teams to trigger compliance processes and obtain approval. These interactions are event-triggered when adapting projects in ways relevant to these teams. Examples of named events were new projects, processing new forms of data, entering new stages of deployment, e.g., from PoC to internal prototype, or adding new features to existing projects. Governance is not generally considered a continuous process after approval is given or a solution is deployed. 

\paragraph{Adversarial Compliance Processes:}  Because compliance functions are rarely directly integrated in development processes, the relationship between compliance teams and project teams was often described as \textit{adversarial}. Projects scrutinized by compliance teams frequently require time-consuming iterative processes to redefine scope and structure before compliance approval is granted. Some interviewees report ``self-censorship'' and reductions in the scope of projects before involving these units.

\paragraph{Limited Employee Capacity:} Interviewees lamented that outside ``digital'' units, employee capacity for digital skills is generally low. Simultaneously, their workload and expected throughput is frequently very high, leading to the perception of additional duties, such as transition to new technologies, oversight of imperfectly functioning digital tools, or development of model evaluation metrics, as burdens rather than reliefs.

\paragraph{Success of Breaking Hierarchy:} Interviewees partaking in models of collaboration that break these  hierarchical and adversarial structures report higher project success and smoother governance integration. Examples include cross-functional expert steering teams that integrate governance functions in project management, and ad-hoc interest networks of AI specialists from many units across an organization.

\section{Discussion} \label{sec:map-challenge}

We find strong evidence for our hypothesis that current governance structures face severe challenges in adapting to agent governance. In particular, our interviews reveal agent governance requirements may produce \textit{intensifications} of governance challenges already familiar to PSOs from existing digitalization projects and implementations of non-agentic AI systems. Here, we first aggregate three classes of shortcomings and conclude by highlighting promising paths forward for both agent architectures and PSO governance structures.

 \begin{enumerate}
     \item \textbf{Continous oversight} is required to translate mechanical visibility into accountability at the operational level. It likely cannot be guaranteed entirely by segmented structures that isolate governance requirements in separated teams, or by processes that are event-triggered: the frequency of events produced by agents exponentiates communication costs between operative and governance units, which are already prohibitive. As corroborated by most interviewees, governance responsibilities must therefore be diffused towards the end-users: the implementing operational departments whose work is augmented by agents.
     \item \textbf{New governance capabilities} are required throughout the agent development cycle. Some specific expert capabilities, such as pre-deployment testing and occasional auditing, may be well covered by new departments within existing governance structures. Visibility and evaluation, however, require much deeper integration between subject knowledge and technical understanding --- again necessitating upskilling and a redefined role for operational workers. This in turn may amplify a variety of risks related to how these workers, who now take on dual roles as overseers, are influenced by the technology in the exercise of their own discretion \cite{de_boer_automation_2023, bullock_2020_ArtificialIntelligenceBureaucraticformdiscretion}. Some degree of central oversight must thus likely remain.
     \item \textbf{Interdepartmental coordination} is required. Agents, which handle processes and connect to tools and databases across traditional departmental boundaries, intensify an existing digital governance challenge frequently named by interviewees: collaboration and responsibility attribution for increasingly complex, cross-cutting projects. Required adaptations include mechanisms for cross-departmental coordination, and sufficient in-department governance competency to avoid dependence on mediation and consultation by external units. An interviewee already involved in structures enabling this, both for governance and for project coordination, cited them as ``unlocking the potential" for many projects.
 \end{enumerate}

Collectively, these developments imply successful agent governance is centrally coordinated, but diffused throughout PSOs. We visualize this transition in figure \ref{fig:governance-structure}. This matrix organization \cite{turk_matrix_2018} is evocative of governance models that iterate on strictly bureaucratic structures, such as those mentioned above (§\ref{sec:theory}). A promising direction of future research is therefore to evaluate which features of these neo-Weberian governance approaches may be promising for agent governance --- especially as the uptake of agentic systems drives more fundamental transformation of state structures \cite{ilves2025agentic}. We propose three design principles for the tailoring of technical work on agent oversight to public-sector contexts.

\begin{enumerate}
\item \textbf{Design observability tooling to be utilized by either technical external teams, or subject matter experts.}

Existing agent observability tooling, including logging of actions and tools, control over autonomy levels, and evaluation metrics, frequently interweaves subject-specific online oversight and technical, abstract offline oversight in combined interfaces and applications \cite{dong_agentops_2024}. Our interviews show that PSOs frequently maintain organizational separation between these groups of tasks (§\ref{sec:governance-pso}). Oversight tooling should therefore be designed with delegation to these distinct groups in mind.

\item \textbf{Design oversight interfaces in collaboration with non-technical public servants where online supervision is required.}

Non-technical public servants are usually ill-suited and not enabled by the organization to translate technical oversight metrics into those required for performance of their duties. Mechanisms by which they control and oversee agent actions must therefore reflect the requirements, logics, and language of the individual end user and the organization, rather than derive from the provided technology (§\ref{sec:gov-agents}). Beyond enabling non-expert oversight, this allows the attribution of responsibility for agent actions to humans.

\item \textbf{Design agent oversight that anticipates interaction with legacy systems.}

While the potential of agents (§\ref{sec:agent-potential}), is enormous in the public sector, the ideal-state implementation relies on the prior end-to-end digitalization of existing services, and is unlikely to be realized in PSOs with fragmented, often partially manual, processes across legacy systems (§\ref{sec:gov-agents}). Agent oversight systems must therefore allow for this variability. This includes designing systems that do not assume complete process observability, and incorporating human-in-the-loop fallback mechanisms for scenarios where automated oversight is insufficient.
 \end{enumerate}


\section*{Limitations}
This study has several limitations that should be acknowledged: \textit{(1) Institutional scope:} Our analysis is limited to German public sector organizations with their specific administrative traditions. While the results are indicative of challenges faced by public sector organisations, they cannot be directly applied to other administrative contexts. \textit{(2) Sample characteristics:} Our sample of six qualitative expert interviews, while providing consistent corroboration of our core hypotheses, may not capture the full spectrum of governance challenges across all public sector organizations. \textit{(3) Implementation Challenges:} Our framework requires context-specific adaptation across diverse institutional structures. Our focus on PSOs as standalone entities does not capture their integration with private-sector institutions for LLM or agent technology, infrastructure, product development, auditing and oversight services.

\section*{Acknowledgements}
JR was supported by the Engineering and Physical Sciences Research Council [Grant Number EP/W524311/1]. JB was supported by the Federal Ministry of Education and Research of Germany [Grant Number 16DII131]. CS and JB acknowledge funding by the Hertie School's program "AI and Data Science for the Public Sector", funded by the Dieter Schwartz Foundation. 

\bibliography{custom,zotero}

\appendix

\section{Interview Guide} \label{app:interview-guide}
\small

\subsection{Research Framework and Methodology}

This guide is grounded in a literature review of LLM agent governance best practices. The interviews will be recorded, transcribed, and stored on secure University servers with access limited to project researchers. All data will be pseudonymized immediately after collection, retaining only minimal demographic information (agency and seniority level). Analysis will utilize coding based on our Agent governance readiness framework to identify gaps and potential, with support from secure LLMs for initial coding. Findings will be reported following COREQ guidelines.

\subsection{Introduction for Interviewer}

This interview guide is designed to assess current AI governance practices in German public sector organizations and identify gaps in preparedness for implementing LLM-based agents. The interview should take approximately 30 minutes. Begin by introducing yourself and providing context on LLM-based agents before proceeding with the questions.

Interviews will be conducted in German by a native German speaker familiar with the public sector context. Focus on creating a comfortable environment for honest discussion, and adapt questions based on the participant's role and familiarity with AI systems.

\subsection{Introduction for Participants}

\textit{[Read to participant in German]}

Thank you for participating in this interview. Today we'll be discussing governance practices for digital technologies and AI in your organization, with a particular focus on how these practices might apply to LLM-based agents.

Your responses will help us understand your organization's governance structures in a neutral and non-opinionated way, focusing on how current governance might need to adapt for the responsible implementation of these technologies. While we are primarily interested in understanding the formal governance processes, we are also keen to hear about your personal experience with these structures, including any challenges or opportunities you have encountered. As such, your personal experience with these processes in your organization are highly valuable. 

This interview will last around 30 minutes. It will be recorded and transcribed for research purposes. All information will be pseudonymized or aggregated in the paper, and only the research team will have access to the transcript. The findings will contribute to developing better governance frameworks for AI agents in public sector organizations.

(If relevant) Lastly, I want to acknowledge explicitly that we know each other from the [certificate program]. While this familiarity may shape some aspects of our discussion, I will be following an interview guide here that does not presuppose this.

Before we begin, let me briefly explain what we mean by "LLM-based agents." As we briefly touched upon in the course, these are AI systems built on large language models that can perform tasks with some degree of autonomy. Unlike simple chatbots that just respond to queries, these agents can:

\begin{itemize}
    \item Execute multi-step processes
    \item Access and utilize various tools and systems
    \item Make certain decisions within defined parameters
    \item Potentially interact with other systems and databases
\end{itemize}
For example, an LLM-based agent might automatically draft responses to citizen queries about building regulations by accessing relevant databases, interpreting regulations, and formulating appropriate responses.

Do you have any questions before we begin?

\subsection{Questions}

\textbf{1. To start, we’d like to gather some basic background information about your role and experience.}
\begin{itemize}
  \item What is your position/role in the organization?
  \item Seniority level (years in current role/public sector)?
  \item Do you have experience with digital transformation and/or AI projects in your organization? If yes, what role do you generally take in them?
\end{itemize}

\textbf{2. Could you describe how NEW digital projects are currently managed in your organization?}
\begin{itemize}
  \item How are decisions about implementing new digital systems made?
  \item Are there dedicated teams or individuals focused on digital project governance?
  \item In what ways, if any, do AI-related projects differ from traditional digital projects in your organization?
\end{itemize}

\textbf{3. When implementing a new digital system or tool, what approval processes or oversight mechanisms are typically involved?}
\begin{itemize}
  \item Where in the organization are these specialized departments located, and how do they interact with other units?
  \item What are the processes for bringing them into a digital project?
  \item At what stage of implementation do they typically get involved?
  \item Are there differences in governance processes, oversight, or approval mechanisms between AI and non-AI projects?
\end{itemize}

\textbf{4. How does coordination work across departments when implementing digital systems that affect multiple units? Consider, for example, projects involving multiple \textit{Fachabteilungen}, or projects implemented by \textit{internal third parties}, such as centres of excellence.}
\begin{itemize}
  \item What challenges have you experienced with cross-departmental digital initiatives?
  \item How are responsibilities divided when a system spans multiple departments?
  \item Are there formal mechanisms for cross-departmental collaboration?
  \item Which governance skills/tasks are where; if you are in a \textit{Fachabteilung}, are they in your team?
\end{itemize}

\textbf{5. What capabilities does your organization have for evaluating the technical performance and safety of AI systems before deployment?}
\begin{itemize}
  \item Who conducts these evaluations?
  \item What metrics or standards are typically used?
  \item Do you have the technical expertise internally, or do you rely on external evaluators?
  \item How do you assess potential risks or failure modes?
\end{itemize}

\textbf{6. How does your organization approach ongoing monitoring and security for AI systems after they've been deployed?}
\begin{itemize}
  \item What mechanisms exist for detecting and responding to potential misuse or system failures?
  \item How are incident response procedures structured?
  \item How often are deployed AI systems reviewed or reassessed?
\end{itemize}

\textbf{7. What procedures exist for documenting and maintaining visibility into how AI systems function and make decisions?}
\begin{itemize}
  \item How is information about AI systems communicated to relevant stakeholders?
  \item Are there requirements for explainability or transparency?
  \item How do you ensure citizens understand when they're interacting with AI systems?
\end{itemize}

\textbf{8. Considering LLM-based agents that can perform tasks with some autonomy, what aspects of your current governance approach do you think would need to change?}
\begin{itemize}
  \item What new challenges do you anticipate?
  \item What resources or capabilities would need to be developed?
  \item How would you approach the question of accountability for agent decisions?
\end{itemize}

\textbf{9. How would you assess your organization's current capacity in terms of staff expertise and resources for governing more autonomous AI systems like LLM-based agents?}
\begin{itemize}
  \item Can you provide examples of any existing governance structures or processes that might already be applicable to autonomous AI agents?
  \item How do you think governance would need to change for AI agents capable of executing multi-step tasks independently, such as processing permit applications or responding to citizen inquiries based on real-time data?
  \item Where in the organization is this expertise currently?
  \item What training or skill development would be needed to manage these new governance requirements?
  \item Are there resource constraints that would affect this capacity?
\end{itemize}

\textbf{10. Is there anything else you'd like to share about AI governance in your organization that we haven't covered?}

Thank you for your time and insights. Your responses will help us better understand how public sector organizations can adapt governance structures for emerging AI agent technologies.

The findings from this research will be analyzed according to our governance readiness framework and may be used to develop recommendations for public sector organizations. We will follow up with a summary of our research findings once the study is complete. If you have any questions or additional thoughts later, please feel free to contact us.

\subsection{Notes for Interviewer}

\begin{itemize}
    \item Adapt questions based on the participant's role and familiarity with AI systems
    \item If time is limited, prioritize questions 1, 2, 3, 8, and 9
    \item For participants with limited AI knowledge, provide additional context as needed
    \item Document any patterns, contradictions, or notable responses for cross-interview analysis
    \item Immediately after the interview, record your observations about non-verbal cues or contextual factors that might be relevant to interpretation
    \item Ensure all recordings are transferred to the secure server within 24 hours
    \item Maintain a research journal documenting methodological decisions and reflections to support rigor
\end{itemize}

\subsection{Post-Interview Protocol}

\begin{enumerate}
    \item Transfer recording to secure University server
    \item Complete interviewer reflection form
    \item Initiate transcription process
    \item Pseudonymize all identifying information
    \item Begin preliminary coding using the governance readiness framework
    \item Document any emerging themes or patterns to explore in subsequent interviews
\end{enumerate}

\section{COREQ checklist} \label{app:coreq}

\begin{description}
\item[\textbf{Domain 1: Research team and reflexivity}]~\\

\textit{Personal Characteristics}
\begin{enumerate}
\item \textbf{Interviewer/facilitator:} Which author/s conducted the interview or focus group?

Answer: All interviews were conducted by the first author.

\item \textbf{Credentials:} What were the researcher's credentials? \textit{(e.g. PhD, MD)}

Answer: MSc.

\item \textbf{Occupation:} What was their occupation at the time of the study?

Answer: Doctoral Researcher / PhD Student.

\item \textbf{Gender:} Was the researcher male or female?

Answer: Male.

\item \textbf{Experience and training:} What experience or training did the researcher have?

Answer: Completed course in qualitative fieldwork and effective interviewing.

\end{enumerate}

\textit{Relationship with participants}
\begin{enumerate}[resume]
\item \textbf{Relationship established:} Was a relationship established prior to study commencement?

Answer: Yes, participants were recruited via a certificate program organized (but not lectured) by the first author.

\item \textbf{Participant knowledge of the interviewer:} What did the participants know about the researcher? \textit{(e.g. personal goals, reasons for doing the research)}

Answer: Outline of research area and interests.

\item \textbf{Interviewer characteristics:} What characteristics were reported about the interviewer/facilitator? \textit{(e.g. bias, assumptions, reasons and interests in the research topic)}

Answer: Assumptions, reasons, and interests were all presented in both the invitation and beginning of each interview (see Appendix \ref{app:interview-guide}).

\end{enumerate}

\item[\textbf{Domain 2: Study design}]~\\

\textit{Theoretical framework}
\begin{enumerate}[resume]
\item \textbf{Methodological orientation and Theory:} What methodological orientation was stated to underpin the study? \textit{(e.g. grounded theory, discourse analysis, ethnography, phenomenology, content analysis)}

Answer: We use content analysis and open-coding grounded in theory of public management to conduct and analyse the interviews.

\end{enumerate}

\textit{Participant selection}
\begin{enumerate}[resume]
\item \textbf{Sampling:} How were participants selected? \textit{(e.g. purposive, convenience, consecutive, snowball)}

Answer: We use purposive sampling to select highly relevant interviewees \cite{palinkasPurposefulSamplingQualitative2015}. Still, there was a degree of convenience sampling as the participants volunteered for interviews.

\item \textbf{Method of approach:} How were participants approached? \textit{(e.g. face-to-face, telephone, mail, email)}?

Answer: We invited for interviews both in a face-to-face class as well as in a follow-up email.

\item \textbf{Sample size:} How many participants were in the study?

Answer: 6 participants.

\item \textbf{Non-participation:} How many people refused to participate or dropped out? Reasons?

Answer: None of the volunteers dropped out of the study. Most participants of the certificate course did not volunteer, often due to information security and privacy reasons.

\end{enumerate}

\textit{Setting}
\begin{enumerate}[resume]
\item \textbf{Setting of data collection:} Where was the data collected? \textit{e.g. home, clinic, workplace}

Answer: The data was collected in auto-transcribed online meetings.

\item \textbf{Presence of non-participants:} Was anyone else present besides the participants and researchers?

Answer: No.

\item \textbf{Description of sample:} What are the important characteristics of the sample? \textit{e.g. demographic data, date}

Answer: All participants sampled worked in bureaucratic institutions for more than a year and were interviewed between February and March 2025. Further characteristics listed in §\ref{sec:methods}.

\end{enumerate}

\textit{Data collection}
\begin{enumerate}[resume]
\item \textbf{Interview guide:} Were questions, prompts, guides provided by the authors? Was it pilot tested?

Answer: We use the interview guide from Appendix \ref{app:interview-guide}. We piloted the questions internally between the authors.

\item \textbf{Repeat interviews:} Were repeat interviews carried out? If yes, how many?

Answer: No.

\item \textbf{Audio/visual recording:} Did the research use audio or visual recording to collect the data?

Answer: We only collected automatic transcripts from the interviews.

\item \textbf{Field notes:} Were field notes made during and/or after the interview or focus group?

Answer: The interviewer wrote informal field notes during and after each interview.

\item \textbf{Duration:} What was the duration of the interviews or focus group?

Answer: Each interview was approximately 30 minutes.

\item \textbf{Data saturation:} Was data saturation discussed?

Answer: Due to the diversity in institutions surveyed and low N, data saturation was not discussed.

\item \textbf{Transcripts returned:} Were transcripts returned to participants for comment and/or correction?

Answer: Each participant was presented with the transcript immediately after the call, and offered the opportunity to comment, redact, or correct. No correction was requested.

\end{enumerate}

\item[\textbf{Domain 3: Analysis and findings}]~\\

\textit{Data analysis}
\begin{enumerate}[resume]
\item \textbf{Number of data coders:} How many data coders coded the data?

Answer: The data were read and analyzed by all authors.

\item \textbf{Description of the coding tree:} Did authors provide a description of the coding tree?

Answer: We use open coding based on the literature review in §\ref{sec:gov-agents}.

\item \textbf{Derivation of themes:} Were themes identified in advance or derived from the data?

Answer: Themes were derived from the data informed by our literature review.

\item \textbf{Software:} What software, if applicable, was used to manage the data?

Answer: No special software on the data.

\item \textbf{Participant checking:} Did participants provide feedback on the findings?

Answer: The findings will be shared in future seminars with the participants. 

\end{enumerate}

\textit{Reporting}
\begin{enumerate}[resume]
\item \textbf{Quotations presented:} Were participant quotations presented to illustrate the themes/findings? Was each quotation identified? \textit{e.g. participant number}

Answer: No, we perform only aggregate analysis to maintain confidentiality.

\item \textbf{Data and findings consistent:} Was there consistency between the data presented and the findings?

Answer: Yes, all presented findings originate from both the literature review and the interviews.

\item \textbf{Clarity of major themes:} Were major themes clearly presented in the findings?

Answer: Yes, see §\ref{sec:governance-pso}.

\item \textbf{Clarity of minor themes:} Is there a description of diverse cases or discussion of minor themes?

Answer: Yes, though elaborate discussion is constrained by the page limit. 

\end{enumerate}
\end{description}

\end{document}